\documentstyle[12pt]{article}
\title{{\hfill {\small Alberta-Thy 10-97}}
\vspace*{.5cm} \\
Plenty of Nothing: Black Hole Entropy in Induced Gravity}
\date{}

\author{V.P. Frolov and D.V. Fursaev\\
\\
{\it Department of Physics, University of Alberta,
Edmonton, Canada T6G 2J1}\\
e-mails: frolov, dfursaev@phys.ualberta.ca}

\begin{document}
\maketitle

\begin{abstract}
We demonstrate how  Sakharov's idea of induced gravity
allows one to explain the statistical-mechanical
origin of the entropy of a black hole.
According  to this idea, gravity becomes dynamical
as the result of quantum effects in the system of
heavy constituents of the underlying theory.
The black hole entropy is related  to the properties of
the vacuum in the induced gravity in the presence of
the horizon. We obtain the Bekenstein-Hawking entropy
by direct counting the states of the constituents.
\end{abstract}

\bigskip\bigskip

\setlength{\unitlength}{1mm}
\baselineskip=.6cm

There are physical phenomena that allow a simple
description but require
tremendous  efforts to
explain them. Without doubts
the origin of black hole entropy is such a problem.
A black hole of mass $M$ radiates as a heated body
\cite{Hawk:75} with temperature
$T_H=(8\pi G M)^{-1}$ and has entropy \cite{Beke:72}
\begin{equation}\label{1}
S^{BH}={{\cal A}\over 4G }  \, ,
\end{equation}
where ${\cal A}$ is the surface area of the
black hole
(${\cal A}=16\pi G^2 M^2$), $G$ is Newton's
constant, and
$c=\hbar=1$.
Statistical mechanics relates entropy to
the measure of disorder and qualitatively it is the
logarithm of the number of
microscopically different states  available for given
values of the
macroscopical parameters.
Are there internal degrees of freedom that are
responsible for the
Bekenstein-Hawking entropy $S^{BH}$? This is the
question that physicists were trying to answer for
almost 25 years.

What makes the problem of black hole entropy so
intriguing? Before discussing
more technical aspects  let us make simple estimations.
Consider for example a
supermassive black hole of $10^9$ solar mass.
According to (\ref{1}), its entropy is about
$ 10^{95}$. This is seven orders of magnitude
larger than the entropy of the
other matter in the visible part of the Universe.
What makes things even more
complicated,
a black hole is simply an empty space-time with a
strong  gravitational field
and ... nothing more. Really, a black hole is
"plenty of nothing", or if we put
this in a more physical way, the phenomenon we
are dealing with is a vacuum in
the strong gravitational field. This conclusion
leaves us practically no other
choice but to try to relate the entropy of the
black hole to  properties of the
physical vacuum in the strong gravitational field.

The black hole entropy is of the same order of
magnitude as the logarithm of
the number of different ways to distribute two signs
$+$ and $-$ over the cells
of Planckian size on the
horizon surface. This estimation suggests that a
reasonable microscopical
explanation of the Bekenstein-Hawking entropy must
be based on the quantum
gravity,
this Holy Grail of the theoretical physics.  The
superstring theory  is the
best what we have and what is often considered as the
modern version of quantum
gravity. Recent observation that  the Bekenstein-Hawking
entropy can be
obtained by  counting of  string (and/or D-brane)
states  is a very interesting
and important result.

Still there are questions. The string calculations
essentially use
supersymmetry and are mainly
restricted to extreme and near-extreme
black holes. Moreover
each  model requires new calculations.
And the last but not the least it remains unclear
why the entropy of a black hole is universal
and does not depend on the details of the theory at
Planckian scales.
Note that the black hole thermodynamics follows from
the low-energy gravitational theory.
That is why one can expect that only a few fundamental
properties of quantum
gravity but not its concrete details are really
important for the
statistical-mechanical explanation of the black
hole entropy.

In the string theory the low-energy gravity with
finite Newton's constant
arises as the collective phenomenon and is the
result of quantum  excitations
of constituents (strings) of the underlying
theory. There is certain similarity
between this mechanism and Sakharov's induced gravity
\cite{Sakharov}. The
low-energy effective action $W[g]$ in the induced
gravity is defined as
a quantum average of the constituent fields
$\Phi$ propagating in a given external
gravitational background $g$
\begin{equation}\label{2}
\exp(-W[g])=\int {\cal D}\Phi\exp(-I[\Phi,g])~~~.
\end{equation}
The Sakharov's basic assumption is that the
gravity becomes dynamical only as
the result of quantum  effects of the constituent
fields.
The gravitons in this picture are analogous to the
phonon field describing collective excitations
of a crystal lattice in the low-temperature limit
of the theory.
Search for the statistical-mechanical origin of
the black hole entropy in Sakharov's approach
\cite{Jacobson}  might help to
understand the
universality of $S^{BH}$.
Here we describe the
mechanism of generation of the Bekenstein-Hawking
entropy in the induced gravity \cite{FFZ,FF}.

Each particular constituent field  in (\ref{2})
gives a divergent contribution to the
effective action $W[g]$.
In the one loop approximation the divergent
terms are local and
of the zero order, linear and quadratic in curvature.
In the induced gravity
the constituents obey additional constraints,
so that the divergences
cancel each other.
It is also assumed that some of the fields
have masses
comparable to the Planck mass and the constraints
are chosen so that the induced cosmological
constant vanishes.
As the result the effective
action $W[g]$ is finite
and in the low-energy limit has the form of
the Einstein-Hilbert action
\begin{equation}\label{3}
W[g]=-{1\over 16\pi G}\left(\int_{\cal M} dV\,
\, R +2\int_{\partial  {\cal
M}}dv \, K\right)+
\ldots \, \,
,
\end{equation}
where Newton's constant $G$ is determined by the
masses of the heavy constituents.
The dots  in (\ref{3}) indicate possible higher
curvature
corrections to $W[g_{\mu\nu}]$
which are suppressed  by the power factors of
$m_i^{-2}$
when the curvature is small. The vacuum Einstein
equations
$\delta W / \delta g^{\mu\nu}=0$ are equivalent to
the requirement that the
vacuum expectation values of the total stress-energy
of the constituents
vanishes
\begin{equation}\label{4}
\langle \hat{T}_{\mu\nu}\rangle
=  0\, .
\end{equation}

The value of the Einstein-Hilbert
action (\ref{3}) calculated on the Gibbons-Hawking
instanton determines the
classical free energy of the black hole, and hence
gives the Bekenstein-Hawking
entropy
$S^{BH}$.  Sakharov's equality (\ref{2}) allows
one to rewrite  the free energy
 as the Euclidean functional integral over
constituent fields on the
Gibbons-Hawking instanton with periodic
(for bosons) and anti-periodic (for
fermions) in the Euclidean time boundary
conditions. In this picture
constituents are thermally excited and
the Bekenstein-Hawking entropy can be
expressed in terms of  the  statistical-mechanical
entropy
\begin{equation}\label{5}
S_R=-\mbox{Tr}~\hat{\rho}\ln\hat{\rho}~~~,~~~
\hat{\rho}={e^{-\hat{H}/T_H} \over
\mbox{Tr}~e^{-\hat{H}/T_H}}~~~.
\end{equation}
Here the operator $\hat{H}$ is the canonical
Hamiltonian of all the constituents.

How can an empty space (vacuum) possess
thermodynamical
properties?
The entropy $S_R$ arises as the result of the
loss of the information about states inside
the black hole horizon
\cite{Sorkin, FN}.
In the induced gravity (entanglement)
entropy (\ref{5})
is calculated for the "heavy"  constituents. This
solves the problems of earlier  attempts to
explain the Bekenstein-Hawking
entropy as the entanglement  entropy of
physical ("light") fields.

Consider the model of induced gravity \cite{FFZ,FF}
that consists of a number
of scalar fields with masses $m_s$ and
a number of Dirac fermions with masses $m_d$.
Scalar fields can have non-minimal couplings and
are described by actions
\begin{equation}\label{6}
I[\phi_s,g]=-\frac 12\int(\phi_s^{,\mu}
\phi_{s,\mu}+m_s^2\phi_s^2+\xi_s
R\phi_s^2)\sqrt{-g}~d^4x~~~.
\end{equation}
The presence of the non-minimally coupled constituents
is important. In this case it is possible to satisfy
the constraints on the parameters $m_s$, $m_d$
and $\xi_s$ which guarantee
the cancellation of the leading ultraviolet
divergencies of the induced gravitational action
$W[g]$, Eq. (\ref{2}).
The induced Newton's constant in this model is
\begin{equation}\label{7}
{1 \over G}=
{1 \over 12\pi} \left(\sum_{s}(1-6\xi_s)~ m_s^2
\ln m_s^2+2\sum_{d}
m^2_d\ln m_d^2\right)~~~.
\end{equation}
The relation between the Bekenstein-Hawking
entropy $S^{BH}$  and the statistical-mechanical
entropy  $S_R$ of the heavy constituents
can be found explicitly and
has the form \cite{FFZ}
\begin{equation}\label{8}
S^{BH}\equiv {1 \over 4G}{\cal A}=S_R-\bar{Q}~~~.
\end{equation}
The important property of $S_R$
is that it diverges because
both fermions and bosons give positive
and infinite
contributions to this
quantity.
An additional term
$\bar{Q}$ in (\ref{8}) is  proportional to the
fluctuations of the
non-minimally coupled scalar  fields
$\hat{\phi}_s$
on the horizon
$\Sigma$ and is the average value  of the
following operator
\begin{equation}\label{9}
\hat{Q}=2\pi\sum_{s}\xi_s \int_{\Sigma}
\hat{\phi}_s^2 \sqrt{\gamma} d^2x~~~.
\end{equation}
The remarkable property of the model is that
for the
same values of the parameters of the constituents,
that guarantee the
finiteness of  $G$, the divergences of $S_R$ are
exactly cancelled by
the divergences of $\bar{Q}$. So in the induced
gravity the
right-hand-side of (\ref{8}) is finite and
reproduces  exactly the  Bekenstein-Hawking
expression.

What is the origin of the compensation mechanism
in (\ref{8}) and what is the
statistical-mechanical meaning of the subtraction
in this relation? The answer
to both questions is in the
properties of the operator $\hat{Q}$.
The operator $\hat{Q}$ is a Noether charge
\cite{Wald:93}
associated with the non-minimal couplings of
the scalar fields.
In statistical-mechanical computations
we consider fields localized
in the black hole exterior.
The charge $\hat{Q}$
determines the difference between the energy
$\hat{E}$
in the external region ${\cal B}$,
defined by means of the stress-energy tensor
and the canonical energy $\hat{H}$
in the same region
\begin{equation}\label{10}
\hat{E}=\int_{\cal B} \hat{T}_{\mu\nu}
\zeta^{\mu}d\sigma^{\nu}=\hat{H}-T_H\hat{Q}\, .
\end{equation}
Here $\zeta^{\mu}$ is the timelike Killing
vector field.

Two energies, $\hat{E}$ and $\hat{H}$,
play essentially different role.
The canonical energy is the value of the Hamiltonian
$\hat{H}$ which is
the
generator of translations of the system
along $\zeta^{\mu}$
and which enters definition (\ref{5})
of the statistical-mechanical entropy $S_R$.
The energy $E$ is the contribution of the
constituents
to the black hole mass. The number
$\nu(E)\Delta E$
of microscopically different physical states
of the constituents in the energy interval
$\Delta E$
near $E=0$ determines the degeneracy of the
black hole mass
spectrum.

Since the Killing
vector $\zeta^\mu$ vanishes at the bifurcation
surface $\Sigma$ of the Killing horizons,
the Hamiltonian $\hat{H}$ is
degenerate. One can add to the system an arbitrary
number of soft modes, i.e. modes with
negligibly small frequencies, without changing
the canonical energy.
On the other hand, only  soft modes contribute
to the
average $\bar{Q}$.
According to (\ref{10}),
this removes the degeneracy of the energy
$\hat{E}$. As the result the infinite
number of thermal states of the
constituents reduces to the finite number
of physical states of the black hole and
$S_R$ reduces
to the Bekenstein-Hawking value.
The
corresponding number density $\nu(E=0)$ of
physical states is
$\exp S^{BH}$ \cite{FF}.
There is a similarity between this mechanism and
gauge theories, soft modes playing the role of
the pure gauge degrees of freedom.

Let us note that the concrete model of the
induced gravity may differ from the
one considered here,
and  may contain, for example,
finite or infinite
number of fields of higher spins. However
our consideration indicates that it
is quite plausible  that the same mechanism
of black-hole entropy generation
still works.

Our discussion can be summarized as follows.
The entropy $S^{BH}$ of a black hole in
induced gravity
is the logarithm  of the number of
different states of
the constituents obeying the condition
$\langle \hat{T}_{\mu\nu}\rangle=0$. The
statistical-mechanical explanation of
the entropy
becomes possible only when one appeals to a more
deep underlying theory in which vacuum is
equipped
with an additional "fine" structure.
The vacuum
in the induced gravity is a ground
state of "heavy"
constituents.
"Light" particles are just long wave collective
excitations over this vacuum. The black
hole entropy
$S^{BH}$
acquires its statistical-mechanical meaning
because of
this additional fine structure of the vacuum of the
underlying theory.
Two important elements that
make statistical-mechanical picture
self-consistent
and universal
are representation of gravity as an induced
phenomenon
and
the ultraviolet finiteness of
the induced gravitational couplings.
These are main lessons the induced gravity
teaches us.

\vspace{12pt}
{\bf Acknowledgements}:\ \ This work was
partially supported  by the Natural
Sciences and Engineering
Research Council of Canada. One of the authors
(V.F.) is
grateful to the Killam Trust for its
financial support.

\end{document}